\documentclass[aps,reprint,noeprint,showpacs,showkeys,superscriptaddress,longbibliography,floatfix]{revtex4-2} %

\usepackage[utf8]{inputenc}
\usepackage{cmap} 
\usepackage[T1]{fontenc}
\usepackage{amsmath,amssymb,mathtools}

\usepackage{hyperref}
\usepackage[dvipsnames]{xcolor}
\hypersetup{
    colorlinks,
    linkcolor={blue!80!black},
    citecolor={blue!80!black},
    urlcolor={blue!80!black}
}

\begin{document}
 
\title{Aging following a zero-temperature quench in the $d=3$ Ising model}
\author{Denis Gessert}
\email{denis.gessert@itp.uni-leipzig.de}
\affiliation{Institut f\"ur Theoretische Physik, Universit\"at Leipzig, IPF 231101, 04081 Leipzig, Germany}
\affiliation{Centre for Fluid and Complex Systems, Coventry University, Coventry CV1~5FB, United Kingdom}
\author{Henrik Christiansen}
\email{henrik.christiansen@itp.uni-leipzig.de}
\altaffiliation{Present address: NEC Laboratories Europe GmbH, Kurfürsten-Anlage 36, 69115 Heidelberg, Germany.}
\affiliation{Institut f\"ur Theoretische Physik, Universit\"at Leipzig, IPF 231101, 04081 Leipzig, Germany}
\author{Wolfhard Janke}
\email{wolfhard.janke@itp.uni-leipzig.de}
\affiliation{Institut f\"ur Theoretische Physik, Universit\"at Leipzig, IPF 231101, 04081 Leipzig, Germany}
\date{\today}
 
\begin{abstract}
  Aging in phase-ordering kinetics of the $d=3$ Ising model following a quench from infinite to zero temperature is studied by means of Monte Carlo simulations. In this model the two-time spin-spin autocorrelator $C_\text{ag}$ is expected to obey dynamical scaling and to follow asymptotically a power-law decay with the autocorrelation exponent $\lambda$. Previous work indicated that the lower Fisher-Huse bound of $\lambda\geq d/2 = 1.5$ is violated in this model. Using much larger systems than previously studied, the instantaneous exponent for $\lambda$ we obtain at late times does \emph{not} disagree with this bound. By conducting systematic fits to the data of $C_\text{ag}$ using different ansaetze for the leading correction term, we find $\lambda = 1.58(14)$ with most of error attributed to the systematic uncertainty regarding the ansaetze. This result is in contrast to the recent report that below the roughening transition universality might be violated.
\end{abstract}
\maketitle

\section{Introduction}

Aging phenomena, quite generally, arise in out-of-equilibrium systems for times shorter than the time it takes to reach a thermodynamically stable state~\cite{puri2009kinetics,henkel2010non}
and are frequently studied in glassy systems~\cite{Henkel2007, Bouchaud1997, bouchaud2000aging}.
Aging in non-glassy systems, such as the paradigmatic Ising model given by the Hamiltonian $\mathcal{H} = - \sum_{\langle i j \rangle} \sigma_i \sigma_j\,$ with the spin variables $\sigma_i \in \{-1,+1\}$, is more accessible by theoretical analysis~\cite{henkel2010non}. For quenches to below the critical temperature $T_c$, most numerical work on aging phenomena in phase ordering processes both in two~\cite{Fisher1988,humayun1991non,Liu1991,Henkel2004,Lorenz2007} and three~\cite{huse1989remanent,Henkel2003,Midya2014} dimensions are in good agreement  with the available theoretical bounds~\cite{Fisher1988}, as well as compatible with some approximate predictions~\cite{ohta1982universal,Liu1991}. One exception are quenches to particularly low temperatures in three dimensions where numerical work so far has been inconclusive~\cite{Fisher1988,das2017kinetics,chakraborty2017coarsening,vadakkayil2019finite,Vadakkayil2022}.

The phase-ordering kinetics following such a quench is described by the characteristic length scale $\ell(t)$ of domains of like spins which is predicted to follow the power law
\begin{equation}
  \ell(t) \propto t^{\alpha}
\end{equation}
with $\alpha = 1/2$ for nearest-neighbor interacting systems, irrespective of the spatial dimension $d$~\cite{bray2002theory}. In the case of a zero-temperature quench in $d=3$ many works~\cite{Fisher1988,shore1992logarithmically,Corberi2008,das2012finite,vadakkayil2019finite}, however, have found the growth to be slower than expected, which was attributed to pre-asymptotic effects~\cite{Corberi2008}. We recently reported results from simulations of much larger systems and found that instead of approaching $1/2$, $\alpha$ actually (at least pre-asymptotically) takes values larger than $1/2$~\cite{Gessert2023}. We then conjectured that at early times the formation of sponge-like structures of the domains hinders the growth while at later time the collapse of said structures does enhance the growth rate.

To better understand this non-equilibrium process we study its aging characteristics by probing the two-time spin-spin autocorrelator given by
\begin{equation}
  C_\text{ag}(t,t_w) = \langle \sigma_i(t) \sigma_i(t_w) \rangle\,, \label{equ:aging}
\end{equation}
where $t$ is the observation time and $t_w$ the waiting time. One expects dynamical scaling in the scaling variable $y = t / t_w$ and that~\cite{puri2009kinetics,henkel2010non}
\begin{equation}
  C_\text{ag}(t,t_w) = f_C(y) \stackrel{y\rightarrow\infty}{\longrightarrow} f_{C,\infty} ~ y^{-\lambda / z} \label{equ:agingT}
\end{equation}
where $f_C$ is the corresponding scaling function, $\lambda$ the autocorrelation exponent, $z = 1/\alpha$ the dynamical exponent, and $f_{C,\infty}$ the amplitude of the asymptotic power law. As usual it is assumed that both $t_w$ and $t-t_w$ are much larger than some microscopic reference time scale $t_\text{micro}$.

The goal of this work is to probe for dynamical scaling and to estimate the autocorrelation exponent $\lambda$ in Eq.~(\ref{equ:agingT}). An early approximation by Ohta \textit{et al.}~\cite{ohta1982universal} yields $\lambda = d/2$~\cite{Yeung1990}, with $d$ being the spatial dimension.
More generally, Fisher and Huse (FH)~\cite{Fisher1988} argued that $\lambda$ has to be in the interval $[d/2,d]$, which corresponds to $\lambda$ inbetween $1.5$ and $3$ here.
Numerically, they find $\lambda$ to effectively vary with time and that it is below or equal to their lower bound using very small systems of $N = 80^3$ spins.
A later approximation by Liu and Mazenko (LM)~\cite{Liu1991} predicted a value of $\lambda\approx 1.67$. More recent numerical work, the most recent being Refs.~\cite{das2017kinetics,chakraborty2017coarsening,vadakkayil2019finite,Vadakkayil2022}, report values of $\lambda$ far below the LM value and outside the FH bound.
As in equilibrium the nature of interfaces changes at the roughening transition, it appears plausible that this may also cause different nonequilibrium behavior.
Specifically, it was argued~\cite{Vadakkayil2022} that these anomalies arise as a result of non-universal behavior below the roughening temperature $T_R\approx 0.54 T_c$~\cite{Hasenbusch1996}, where in equilibrium interfaces are locally flat instead of rough as they would be for $T > T_R$~\cite{Beijeren1987}. This effect is most dominantly seen at zero temperature, which is the case considered here.
These authors already pointed out the need for considering large systems for studying dynamics at such low temperatures~\cite{das2017kinetics}. We believe that the aging results for their considered box lengths are still affected by transients which might have led to such conclusion. We revisit this complex problem, armed with resources and methods that are capable of handling much bigger systems over long periods.

The rest of the paper is organized as follows. In Sec.~\ref{sec:methods} we describe the used methodology. Our numerical results are presented in Sec.~\ref{sec:results} and we conclude in Sec.~\ref{sec:conclusion}.

\section{Methods} 
\label{sec:methods}

Part of the later analysis invokes direct use of the characteristic length scale $\ell(t)$
which we obtain by measuring the two-point equal-time correlation function $C(r,t) = \langle \sigma_i \sigma_j \rangle - \langle \sigma_i \rangle \langle \sigma_j \rangle$. As $C(r,t)$ obeys dynamical scaling, i.e., $C(r,t) = \tilde{C} (r/\ell(t))$, the length $\ell(t)$ can be extracted by finding the distance $r = \ell(t)$ at which the correlation function has decayed to 50\% (or any other percentage; for details see Sec.~III of the Supplementary Material in Ref.~\cite{Gessert2023}). 

The simulations were carried out for non-conserved order parameter using Monte Carlo (MC) simulations utilizing the Glauber acceptance criterion~\cite{glauber1963time}, which at zero temperature translates to accepting (rejecting) every move that decreases (increases) the internal energy of the system, and accepting a move without a change in energy with 50\% chance. As at low temperatures most moves would otherwise be rejected, we use the rejection-free $n$-fold way update~\cite{n-fold}, which has identical dynamics as the Glauber update and speeds up the simulations significantly.

All data presented were averaged over 40 independent realizations for the system sizes $L=512, 1024$, and over 36 realizations for $L=1536$. We make use of periodic boundary conditions. The simulation for each realization is carried out by using a different initial random configuration (corresponding to $T=\infty$ with close-to-zero magnetization per lattice site, i.e., $m\approx 0$) and using a different random number seed for each time evolution. Throughout, error bars correspond to the standard error obtained by using the Jackknife method~\cite{efron:82}.

\section{Results}
\label{sec:results}
\begin{figure}
  \centering
  \includegraphics[width=0.45\textwidth]{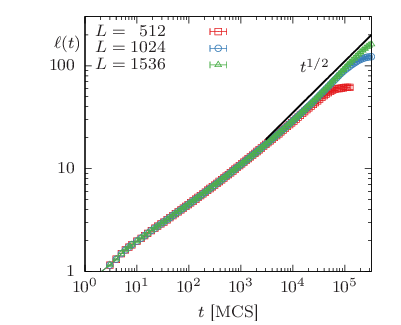}
  \caption{Characteristic length scale $\ell$ vs.\ MC time $t$. Although power-law growth with exponent $1/2$ (solid line) is expected, at early times slower and at late times faster growth is observed.}
  \label{fig:ell}
\end{figure}

Figure~\ref{fig:ell} shows the characteristic length scale as a function of time $t$ in Monte Carlo sweeps (MCS). Although roughly compatible with the expected power-law growth $\propto t^{1/2}$ (solid line), the data shows some deviations: At early times it grows much slower and for $L=1536$ at late times it grows faster. For a detailed analysis see Ref.~\cite{Gessert2023} where we also present the instantaneous growth exponent which shows the growth faster than $t^{1/2}$ very clearly.

One problem in the estimation of $\lambda$ is that when plotting $C_\text{ag}(t,t_w)$ as a function of $t / t_w$ and fitting a power law one obtains only the exponent ratio $\lambda/z$. As outlined above, also the value for $z = 1/\alpha$ is still debated and effectively changes with time which complicates the estimation of $\lambda$. To avoid this problem, it is common~\cite{das2017kinetics,chakraborty2017coarsening, vadakkayil2019finite,Vadakkayil2022} to choose $x = \ell(t) / \ell_w$ as a scaling variable where $\ell_w = \ell(t_w)$, giving rise to the aging relation 
\begin{equation}
  C_\text{ag}(t,t_w) = \tilde{f}_C(x) \stackrel{x\rightarrow\infty}{\longrightarrow} \tilde{f}_{C,\infty} ~ x^{-\lambda}. \label{equ:agingEll}
\end{equation}

\begin{figure}
  \centering
  \includegraphics[width=0.45\textwidth]{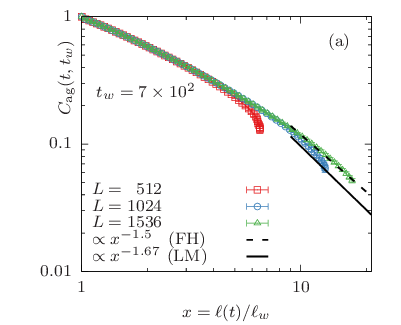}
  \includegraphics[width=0.45\textwidth]{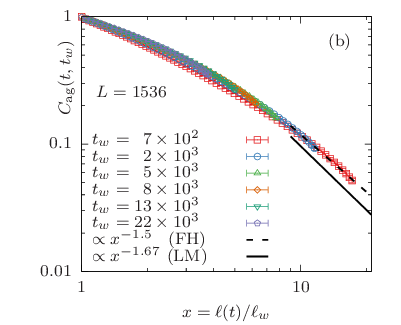}
  \caption{Autocorrelation function $C_\text{ag}(t,t_w)$ plotted against $\ell(t)/\ell_w$. The dashed lines show the FH lower bound~\cite{Fisher1988} and the solid lines the LM value~\cite{Liu1991}. (a) Results for different system sizes $L$ and (b) for different waiting times $t_w$ in units of MCS. The actual values for the mnemonic $t_w$ shown here and in subsequent figures are provided in Appendix~\ref{app:actualWaitingTimes}.}
  \label{fig:aging_ell}
\end{figure}
 
In Fig.~\ref{fig:aging_ell} the autocorrelation function $C_\text{ag}$ is plotted vs.\ $x = \ell(t)/\ell_w$ for (a) different system sizes and (b) different waiting times (cf.\ Appendix~\ref{app:actualWaitingTimes}).
As can be seen in Fig.~\ref{fig:aging_ell}(a), finite-size effects occur in the usual way: For small $x$ the data points fall on a common curve and for larger $x$ the data sets bend away from this common curve in order of ascending system size.
Dynamical scaling is probed by considering the curves for different $t_w$ shown in Fig.~\ref{fig:aging_ell}(b). These collapse as expected from Eq.~(\ref{equ:agingEll}), although particularly for small $t_w$ and for small $x$ not perfectly. 
Besides the suboptimal dynamical scaling, $C_\text{ag}$ also does not (yet) decay according to a pure power law. 
For $x \gtrsim 10$ it can be seen that the function decays at least as fast as the FH lower bound and is not incompatible with the LM value. This leads us to believe that the FH bound is not violated (unlike previously conjectured).%

\begin{figure}
  \centering
  \includegraphics[width=0.45\textwidth]{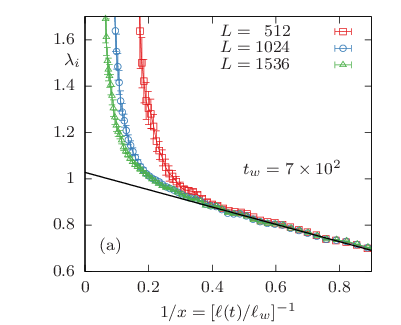}
  \includegraphics[width=0.45\textwidth]{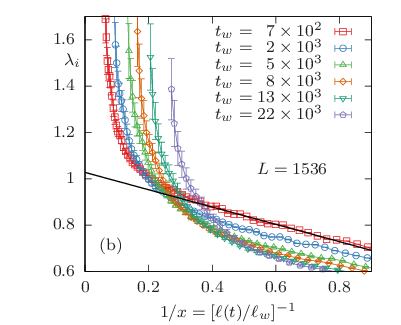}

  \caption{Instantaneous autocorrelation exponent $\lambda_i(t,t_w)$ vs.\ $[\ell(t)/\ell_w]^{-1}$ (a) for different system sizes and (b) for different waiting times $t_w$. The solid black lines in (a) and (b) show an extrapolation in $1/x$ as performed in Ref.~\cite{vadakkayil2019finite}.}
  \label{fig:aging_lambdaI}
\end{figure}

To study the effective variation of the power-law exponent for $C_\text{ag}$ with time more quantitatively, one can consider the instantaneous autocorrelation exponent 
\begin{equation}
  \lambda_i = - \frac{\mathrm{d} \ln C_\text{ag}}{\mathrm{d} \ln x}. \label{equ:lambdaI}
\end{equation}
The results for $\lambda_i$ are shown in Fig.~\ref{fig:aging_lambdaI} as a function of $1/x$. As can be seen, $\lambda_i$ for $1/x > 0.4$ is practically a linear function which seemingly allows a linear extrapolation to $1/x \rightarrow 0$ (solid line). Vadakkayil \textit{et al.}~\cite{vadakkayil2019finite} have performed such extrapolation using system sizes up to $512^3$ and found that $\lambda_i$ approaches a value of $\approx 1.2$ which is well below the FH lower bound~\cite{Fisher1988} and far below the LM value~\cite{Liu1991} of $1.67$.
When performing this exercise we obtain a similar value for $\lambda$, viz.\ $\lambda \approx 1.0$. This somewhat smaller value than that in Ref.~\cite{vadakkayil2019finite} may be due to $\ell$ having been measured differently~\footnote{Reference~\cite{vadakkayil2019finite} defines $\ell(t)$ as the first moment of the domain-size distribution, whereas we obtain $\ell(t)$ through $C(r,t)$.}. Only by studying larger systems, it becomes apparent that the linear behavior is in fact misleading: 
The data for $L=1024$ and $L=1536$ in Fig.~\ref{fig:aging_lambdaI}(a) on the interval $1/x \in [0.2,0.4]$ are still in good agreement and well above the extrapolating line.
Only at $1/x \lesssim 0.2$ the data for $L=1024$ leaves the common curve, which marks the onset of finite-size effects for this system size.
Therefore, on this interval finite-size effects are negligible and the extrapolating line does not provide any information on the true asymptotic behavior.
The key problem, it appears, is that finite-size effects in $\lambda_i$ show themselves in form of the curve drifting towards higher $\lambda_i$.
As a consequence, an increase in $\lambda_i$ (before the onset of finite-size effects) may easily be mistaken for a finite-size effect, and this indeed seems to be the case.
What is more, from Fig.~\ref{fig:aging_lambdaI}(b) it becomes immediately clear that the value obtained from the extrapolation strongly depends upon the choice of $t_w$. It shows $\lambda_i$ for various waiting times $t_w$ and for each $t_w$ we would get a different estimate for the asymptotic $\lambda$ if we were to carry out the extrapolation.

The reason for finite-size effects causing $\lambda_i$ to shoot up is that $\ell(t)$ stagnates at some value when finite-size effects set in, and thus any further decreasing autocorrelation will effectively have the same $x$-value when plotted against $\ell/\ell_w$ as can be seen in Fig.~\ref{fig:aging_ell}(a). Consequentially, this gives rise to very steep slopes~$\lambda_i$ visible in Fig.~\ref{fig:aging_lambdaI}. Hence, it is worth considering $C_\text{ag}$ as a function of $\sqrt{y} \equiv \sqrt{t / t_w}$ (corresponding to the expected asymptotic behavior of $\ell(t)/\ell_w$ for an infinite system). In this case finite-size effects result in smaller (absolute) slopes which allows to differentiate an increase in the analogously to Eq.~(\ref{equ:lambdaI}) defined instantaneous exponent
\begin{equation}
  [\lambda / z]_i \times 2= -\frac{\mathrm{d} \ln C_\text{ag}}{\mathrm{d} \ln \sqrt{y}} \label{equ:lambdaZi}
\end{equation}
from finite-size effects.

\begin{figure}
  \centering
  \includegraphics[width=0.45\textwidth]{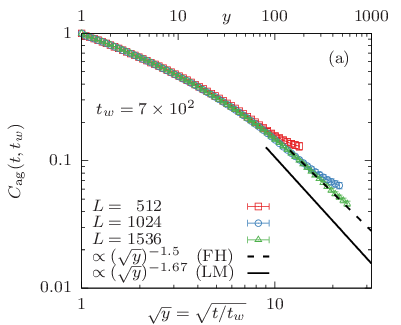}
  \includegraphics[width=0.45\textwidth]{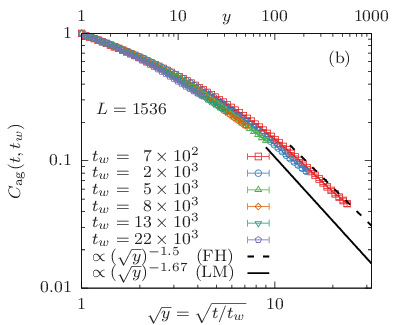}
  \caption{As Fig.~\ref{fig:aging_ell} but using $\sqrt{y}$ as scaling variable instead: Autocorrelation function $C_\text{ag}(t,t_w)$ plotted against $\sqrt{y}$. The dashed lines show the FH lower bound~\cite{Fisher1988} and solid lines the LM value~\cite{Liu1991}. (a) Results for different system sizes $L$ and (b) for different waiting times $t_w$. In both panels the upper abscissa shows $y$ (as opposed to $\sqrt{y}$ in the lower abscissa).}
  \label{fig:aging_t}
\end{figure}

We thus repeat the same type of analysis as above but using $\sqrt{y}$ as the scaling variable instead of $x$. In Fig.~\ref{fig:aging_t} we show the two-time autocorrelator (a) for different system sizes and (b) for different waiting times. As outlined above, finite-size effects appear in form of a reduced decay of $C_\text{ag}$ when plotted against $\sqrt{y}$; see Fig.~\ref{fig:aging_t}(a). Similar to Fig.~\ref{fig:aging_ell}(b), the quality of the collapse of $C_\text{ag}$ for different waiting times [Fig.~\ref{fig:aging_t}(b)] is moderate whereas $C_\text{ag}$ for different $L$ agrees well before the onset of finite-size effects.

\begin{figure}
  \centering
  \includegraphics[width=0.45\textwidth]{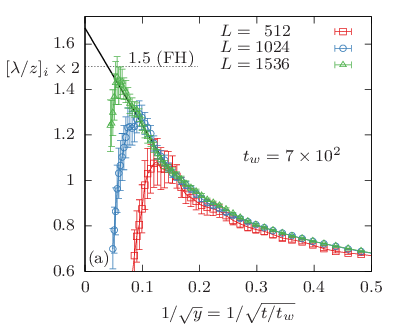}
  \includegraphics[width=0.45\textwidth]{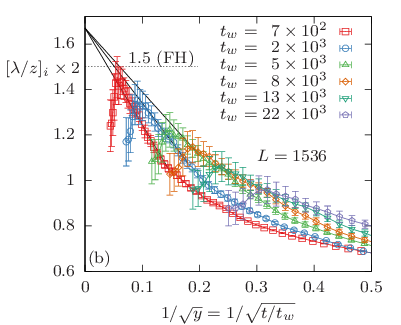}
  \caption{As Fig.~\ref{fig:aging_lambdaI} but using $\sqrt{y}$ as scaling variable instead: Instantaneous exponent $[\lambda/z]_i \times 2$ vs.\ $(t/t_w)^{-1/2}$ (a) for different system sizes and (b) for different waiting times $t_w$. In both cases the solid lines are guide to the eye connecting the data points to the LM value as $1/\sqrt{y} \rightarrow 0$. The value of the FH lower bound $1.5$ is shown with dotted lines.}
  \label{fig:aging_lambdaZi}
\end{figure}

In Fig.~\ref{fig:aging_lambdaZi} the instantaneous exponent $[\lambda / z]_i\times 2$ is presented (a) for the studied system sizes and (b) for different waiting times.
As before, it increases at later times (that is, small $1/\sqrt{y}$) which now is clearly distinguished from the finite-size effects (cf.\ Fig.~\ref{fig:aging_lambdaI}), in which the exponent strongly decreases. For $t_w=7 \times 10^2$ in units of MCS [Fig.~\ref{fig:aging_lambdaZi}(a)] alongside the linear behavior for large $1/\sqrt{y}$, now also for small values of the abscissa the exponent changes linearly. The rate at which the exponent changes is not incompatible with the LM value (solid line).

Similar observations can be made when considering larger waiting times [Fig.~\ref{fig:aging_lambdaZi}(b)]: The dataset for ${t_w=2 \times 10^3}$, just before the onset of finite-size effects, follows a straight line with a different slope also consistent with the LM value (solid line). The data for $t_w = 5 \times 10^3$ is not as conclusive
but also compatible with the estimate and for larger waiting times
too little range of the abscissa is covered before finite-size effects set in.

\begin{figure}
  \centering
  \includegraphics[width=0.45\textwidth]{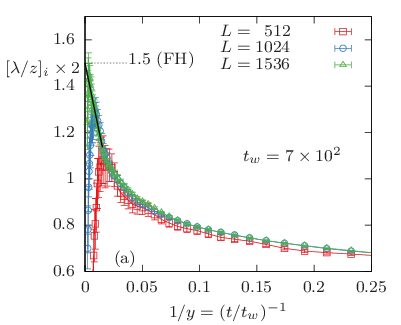}
  \includegraphics[width=0.45\textwidth]{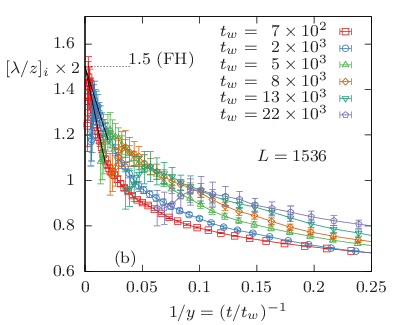}
  \caption{As Fig.~\ref{fig:aging_lambdaZi} but using $1/y$ as abscissa instead of $1/\sqrt{y}$: Instantaneous exponent $[\lambda/z]_i \times 2$ vs.\ $(t/t_w)^{-1}$ (a) for different system sizes and (b) for different waiting times $t_w$. In both cases the solid lines are guide to the eye connecting the data points to the FH lower bound $1.5$ (shown as dotted lines) as $1/y \rightarrow 0$.}
  \label{fig:aging_lambdaZi_y}
\end{figure}

Theoretically, on the other hand, using local-scale invariance (LSI)~\cite{henkel2010non,Henkel2017} and studying exactly solvable models such as the one-dimensional Ising model and the spherical model, one expects the leading correction in $C_\text{ag}$ to be in $1/y$~\cite{Christiansen2020}.
Thus, it seems natural to consider the instantaneous exponent of $C_\text{ag}$ with respect to $y$. 
Of course, log-log plots of $C_{\rm ag}$
against $y$ look precisely the same as in Fig.~\ref{fig:aging_t} with the $y$-values read off from the upper abscissa.
Surprisingly, when carrying out the extrapolation of the instantaneous exponent in $1/y$ (see Fig.~\ref{fig:aging_lambdaZi_y}) we get quite different extrapolating values of similar quality.
Graphically the data appears just as compatible with a linear extrapolation in $1/y$ as it does with an extrapolation in $1/\sqrt{y}$. In this case the intercept was fixed to $1.5$ (the FH lower bound) instead of $1.67$ (the LM value).

Above analysis strongly relies on the use of the instantaneous exponents of $C_\text{ag}$ and the extrapolation in $1/x$, $1/\sqrt{y}$, and $1/y$ in these.
The calculation of the instantaneous exponents requires numerical derivatives subject to various subtleties such as the used stencil and the point density and typically have larger errors than the original data.
We therefore now turn to studying the original data for $C_\text{ag}$ and using correction-to-scaling fits to extract a quantitative estimate for the asymptotic value for $\lambda$.
Note that for this quantitative exercise we are quoting the actual waiting times instead of the mnemonic ones (cf.\ Appendix~\ref{app:actualWaitingTimes}).
The linear behavior with respect to $1/\sqrt{y}$ seen in Fig.~\ref{fig:aging_lambdaZi} suggests empirically that the leading correction term for $C_\text{ag}$ is in $1/\sqrt{y}$, i.e.,
\begin{subequations}
  \begin{equation}
    C_\text{ag} (y) \simeq a y^{-\lambda/z} \left(1 - \frac{c}{\sqrt{y}}\right),\label{equ:CagFitSqrt}
  \end{equation}
  where $a$ is the amplitude of the leading term and $c$ the strength of the first correction term.
  Plugging this into Eq.~(\ref{equ:lambdaZi}) yields
  \begin{equation}
    [\lambda/z]_i \times 2 = \frac {2}{z}\lambda - \frac{c}{\sqrt{y}-c} \stackrel{y\rightarrow\infty}{\longrightarrow} \frac {2}{z} \lambda - \frac{c}{\sqrt{y}}.\label{equ:instaExpSqrt}
  \end{equation}
\end{subequations}
Analogously, a leading correction of $C_\text{ag}$ in $1/y$~\cite{Christiansen2020}, i.e.,
\begin{subequations}
  \begin{equation}
    C_\text{ag} (y) \simeq a y^{-\lambda/z} \left(1 - \frac{c}{y}\right),\label{equ:CagFitLin}
  \end{equation}
  yields
  \begin{equation}
    [\lambda/z]_i \times 2 = \frac {2 }{z}\lambda - \frac{2c}{y-c} \stackrel{y\rightarrow\infty}{\longrightarrow} \frac {2} {z}\lambda - \frac{2c}{y}.\label{equ:instaExpLin}
  \end{equation}
\end{subequations}

\begin{figure*}
  \includegraphics[width=\textwidth]{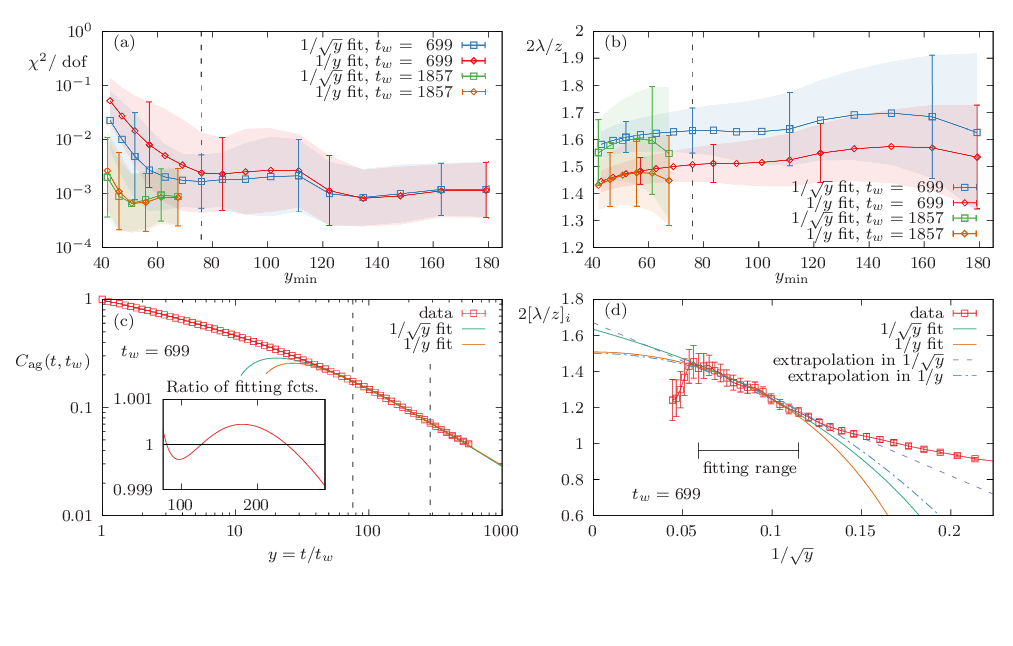}
  \vspace*{-1.5cm}
  \caption{Correction-to-scaling fits on $C_\text{ag}(t,t_w)$ using Eqs.~(\ref{equ:CagFitSqrt}) and (\ref{equ:CagFitLin}), denoted by ``$1/\sqrt{y}$ fit'' and ``$1/y$ fit'', respectively, for $L=1536$. (a) $\chi^2$ per degree of freedom using $t_w=699$ and $1857$ as a function of $y_\text{min}$ with $y_\text{max} = 2 \times 10^5 / t_w$. Colored shades correspond to error bars obtained by Jackknifing~\cite{efron:82}. The dashed vertical line shows the $y_\text{min}$ chosen for subpanels (c) and (d). (b) Numerical value for $2\lambda/z$ obtained from the same fits as in (a). (c) Data and fitting functions for $C_\text{ag}(t,t_w)$ and $t_w=699$. The fitting range $y\in[76,290]$ is indicated by dashed vertical lines. The inset shows that the ratio Eq.~(\ref{equ:CagFitLin})/Eq.~(\ref{equ:CagFitSqrt}) of the two fitting functions varies by less than 0.1\% within the fitting range. (d) Instantaneous exponent as in Fig.~\ref{fig:aging_lambdaZi}. Solid lines show log-derivative of the fits from Eqs.~(\ref{equ:CagFitSqrt}) and (\ref{equ:CagFitLin}), dashed lines are the graphical extrapolations in $1/\sqrt{y}$ and $1/y$ as shown as guide to the eye in Figs.~\ref{fig:aging_lambdaZi} and \ref{fig:aging_lambdaZi_y}, respectively.}
  \label{fig:aging_fits}
\end{figure*}

The results from performing fits of $C_\text{ag}$ with both correction-to-scaling ansaetze are summarized in Fig.~\ref{fig:aging_fits} for $L=1536$. The range of the fits $[y_\text{min},y_\text{max}]$ was determined in the following way: $y_\text{max}$ was fixed as ${2\times 10^5} / t_w$ where $t_\text{max}=2\times 10^5$ is the time when finite-size effects set in, determined through the maximum location of the instantaneous exponent. ($t_\text{max}$ is practically independent of $t_w$.) The lower bound $y_\text{min}$ we have systematically studied by monitoring the goodness of the fit~\footnote{For $y_\text{max}$ the goodness of the fit is not a suitable criterion because of rapidly increasing error bars of $C_\text{ag}$ for large $y$.}; see Fig.~\ref{fig:aging_fits}(a). Due to the strong correlation of the obtained time series, the numerical value obtained for $\chi^2$ as a measure for the goodness of the fit has little significance. However, it is still useful to compare it for different $y_\text{min}$ as a rapidly growing $\chi^2$ for small $y_\text{min}$ can indicate the $y$ at which only a first-order correction is no longer sufficient. 
By doing so, we have found that for $t_w=699$ both fit ansaetze are justified down to $y_\text{min} \approx 65 - 75$. Note that this $y_\text{min}$ is in principle independent of $t_w$. %
Although the $\chi^2$ values obtained for the $1/\sqrt{y}$ fits are slightly smaller than the ones obtained for the $1/y$ fits, both are within error bars and we cannot make any quantitative statement on which ansatz performs better.
The error bars on the fitting parameters were obtained by using the Jackknife method~\cite{efron:82}, that is carrying out many fits with subsets of the data, and then calculating the spread of the obtained parameters.

The estimates of the exponent $2\lambda/z$ obtained from this analysis are shown in Fig.~\ref{fig:aging_fits}(b). The values of the exponent for $t_w=699$ and $1857$ agree until $y_\text{min}\approx60$ which is when there are too few data points~\footnote{For both waiting times the largest $y_\text{min}$ presented corresponds to a fitting range consisting of six data points.} for the larger waiting time which is reflected in quickly growing error bars and a decay in the exponent. For $t_w = 699$ a much larger range of $y_\text{min}$ can be studied. 
Throughout this range the obtained values for $2\lambda/z$ are constant within error bars although a trend of a slightly increasing $2\lambda/z$ with increasing $y_\text{min}$ cannot be ruled out.
The largest systematic uncertainty comes from the choice of the scaling correction.
Throughout, we obtain larger values for $\lambda/z$, when using $\sqrt{y}$ as scaling variable; the reason for this is that $1/\sqrt{y}$ is a stronger correcting term than $1/y$.
Numerically, it is unfortunately impossible to discern which fitting ansatz is more appropriate.

This is demonstrated in Fig.~\ref{fig:aging_fits}(c) where the fitting functions for both ansaetze are shown for the fitting range $y\in[76,290]$.
For the $1/\sqrt{y}$ ansatz using Eq.~(\ref{equ:CagFitSqrt}) we obtain the fitting parameters
\begin{equation}
  2\lambda/z = 1.633(83), \, a = 8.8(2.3), \, c^2 = 7.8(2.5), \label{equ:fittingParamsSqrt}
\end{equation}
and for the $1/y$ ansatz using Eq.~(\ref{equ:CagFitLin})
\begin{equation}
  2\lambda/z = 1.507(64), \, a = 5.35(94), \, c = 11.5(2.5) \label{equ:fittingParamsLin}
\end{equation}
is found.
Note, that we present $c^2$ for the $1/\sqrt{y}$ ansatz, as then in both cases the third parameter is in the same units as $y$, thus allowing straightforward comparison of the correction amplitudes.
At the scale of the plot both fits are indistinguishable from the data except for small $y$ outside the fitting range. 
The inset shows the ratio of the two fitting functions within the fitting interval which varies around $1.0\pm 0.001$. The relative error of the data in the relevant range is about one order of magnitude larger, i.e., $1\% -2 \%$, which leads us to conclude that (at least numerically) we cannot draw any conclusion on which form describes the data better. In Appendix~\ref{app:furtherFittingAnsaetze} we consider yet two other forms for $C_\text{ag}$ for which the asymptotic limit of the instantaneous exponent in Eqs.~(\ref{equ:instaExpSqrt}) and (\ref{equ:instaExpLin}) is exact.

Figure~\ref{fig:aging_fits}(d) shows the instantaneous exponent from Fig.~\ref{fig:aging_lambdaZi} alongside the curves calculated from Eqs.~(\ref{equ:instaExpSqrt}) and (\ref{equ:instaExpLin}) using the fitting parameters from Fig.~\ref{fig:aging_fits}(c). Within the range of the fits resp.~the extrapolations, all lines agree with the data well within error bars. As for the asymptotic value of $2\lambda/z$ they range from $1.5$ to $1.7$ (using the theory value $z=2$) which includes both the FH lower bound of $1.5$ as well as the LM value of $1.67$. As can be seen from Eqs.~(\ref{equ:instaExpSqrt}) and (\ref{equ:instaExpLin}), for $y \rightarrow \infty$ the instantaneous exponents obtained from fits of $C_\text{ag}$ should correspond to the linear extrapolations in the numerical data for the instantaneous exponent.
Because the graphical extrapolations do not take into account the subtracting term in the denominators of Eqs.~(\ref{equ:instaExpSqrt}) and (\ref{equ:instaExpLin}), 
they
deviate slightly from the log-derivatives of the fits.%

This fitting exercise leads to the conservative estimate of $\lambda \in [1.44,1.72]$ (obtained from the upper bound of Eq.~(\ref{equ:fittingParamsSqrt}) and the lower bound of Eq.~(\ref{equ:fittingParamsLin}) assuming that asymptotically $z=2$ holds) which includes the FH lower bound, the value of $1.6$ found at temperatures above $T_R$~\cite{huse1989remanent,Henkel2003,Midya2014}, as well as the LM approximation while excluding low values of $\approx 1.1$ from Refs.~\cite{chakraborty2017coarsening,Vadakkayil2022} and $\approx 1.2$ found in Ref.~\cite{vadakkayil2019finite}.

\section{Conclusion}
\label{sec:conclusion}
We have performed large-scale Monte Carlo simulations with Glauber updates of the three-dimensional Ising model when quenched to $T=0$ for system sizes up to $1536^3$. In this setting we have studied the two-time spin-spin autocorrelation function to probe aging phenomena in this system. The main objective was to test for dynamical scaling and to determine the autocorrelation exponent $\lambda$.

The observed dynamical scaling in our data is not optimal, i.e., the autocorrelator $C_\text{ag}$ does not collapse perfectly when plotted as a function of $t/t_w$ for different waiting times $t_w$. We suspect this to be due to the presence of pre-asymptotic effects even for very large systems and at considerably late times. This is in agreement with late pre-asymptotic effects observed in the characteristic length scale we recently reported on~\cite{Gessert2023}.
Nonetheless, our observations suggest that the FH lower bound is obeyed in this model, which is in contrast to earlier reports~\cite{Fisher1988,das2017kinetics,chakraborty2017coarsening,vadakkayil2019finite,Vadakkayil2022}. At the latest times we could study, the decay of the autocorrelation function is at least as fast as the FH lower bound. As the instantaneous exponent throughout is increasing, an asymptotic value for $\lambda$ below $1.5$ seems very unlikely. In fact, based on our results, even larger values, potentially compatible with a value of $1.6$ as observed in quenches to non-zero temperatures $T > T_R$~\cite{huse1989remanent,Henkel2003,Midya2014}, or with $1.67$ as suggested by Liu and Mazenko~\cite{Liu1991} appear realistic. 

In an attempt to get an as quantitative as possible estimate of the asymptotic value of $\lambda$ from our data, we have carried out correction-to-scaling fits to the two-time spin-spin autocorrelation function data varying both the fitting range $[y_\text{min},y_\text{max}]$ and the waiting time $t_w$. By doing so we obtain a conservative estimate $\lambda = 1.58(14)$ which includes the FH lower bound, the value seen for temperatures above the roughening transition $T_R$ and the LM value. We find larger exponents when using $\sqrt{y}$ as scaling variable than when using $y$. This trend is not specific to using fits of $C_\text{ag}$ but also occurs when carrying out extrapolations in the instantaneous exponents with respect to $1/x = (\ell(t)/\ell_w)^{-1} \simeq (\sqrt{t}/\sqrt{t_w})^{-1} = 1/\sqrt{y}$ as compared to the ones performed with respect to $1/y$ (see Appendix~\ref{app:furtherFittingAnsaetze} for quantitative fits of the instantaneous exponent). For a conclusive answer, however, even much larger systems would need to be studied.

Regarding the recent report of non-universality in quenches to below the roughening transition temperature $T_R$~\cite{Vadakkayil2022}, our results are in favor of universality even for quenches to zero temperature far below $T_R$. An interpretation of the observations in Ref.~\cite{Vadakkayil2022} could be that instead of non-universality for temperatures below $T_R$ strong corrections to scaling are induced that result in an extraordinarily long pre-asymptotic regime, which we expect to be strongest at $T=0$. 
As for temperatures below $T_R$ nonequilibrium interfaces are locally flat for increasing distance with decreasing temperature, it is plausible that pre-asymptotic behavior is seen when the coarsening length-scale is less than the distance at which interfaces are flat. This is consistent with above corrections to scaling strongest at $T=0$ and links the roughening transition to the observed long transients. %
It would be interesting to see whether one makes similar observations for non-zero quench temperatures below $T_R$ when studying systems as large as we did here. In particular, it might be insightful to carry out correction-to-scaling fits for various quench temperatures and monitor the magnitude of the amplitude of the corrections, $c$.

\begin{acknowledgments}
We thank Subir K.\ Das for useful comments on the manuscript.
This project was funded by the Deutsche Forschungsgemeinschaft (DFG, German Research Foundation) under project No.\ 189\,853\,844 -- SFB/TRR 102 (project B04) and the Deutsch-Franz\"osische Hochschule (DFH-UFA) through the Doctoral College ``$\mathbb{L}^4$'' under Grant No.\ CDFA-02-07. We further acknowledge support by the Leipzig Graduate School of Natural Sciences ``BuildMoNa''.

Calculations were performed using the Sulis Tier 2 HPC platform hosted by the Scientific Computing Research Technology Platform at the University of Warwick. Sulis is funded by EPSRC Grant EP/T022108/1 and the HPC Midlands+ consortium. Moreover, we acknowledge the provision of computing time on the parallel computer cluster \emph{Zeus} at Coventry University.
\end{acknowledgments}

\appendix
\section{Actual waiting times}
\label{app:actualWaitingTimes}
\begin{table}[b]
  \caption{Actual waiting times $t_w$ and their rounded mnemonic values usually referred to in the text.}
  \label{tab:actualWaitingTimes}
\begin{tabular}{rc}
  \hline\hline
  ~ actual $t_w$ ~&~ mnemonic $t_w$ ~\\
 \hline
  699  & $\,\ 7\times 10^2$\\
   1857 & $\,\ 2\times 10^3$\\
   4864 & $\,\ 5\times 10^3$\\
   7855 & $\,\ 8\times 10^3$\\
   12674 & $13\times 10^3$\\
   22484 & $22\times 10^3$\\
   \hline\hline
\end{tabular}
\end{table}
The measuring times are evenly spaced on a log scale and we have selected the studied waiting times as a subset of these. Because of this design choice, both the measuring times as well as the waiting times are not round numbers. Additionally, as we chose to use for $t_w$ every tenth point in time for $t < 5000$ and every fifth for $t > 5000$, this translates to quite odd waiting times which is why we have usually referred to their rounded values, except when discussing the quantitative fitting analysis. In Table~\ref{tab:actualWaitingTimes} we show the actual times at which we have saved the spin configurations.

\section{Further fitting ansaetze for the autocorrelation function}
\label{app:furtherFittingAnsaetze}
Previous work such as Refs.~\cite{Midya2014,chakraborty2017coarsening} also used an exponential correction ansatz for $C_\text{ag}$, i.e., 
\begin{subequations}
  \begin{equation}
    C_\text{ag} (y) \simeq a y^{-\lambda/z} \exp \left(-\frac{c}{\sqrt{y}}\right),\label{equ:CagFitExpSqrtY}
  \end{equation}
  as it yields an instantaneous exponent linear in $1/\sqrt{y}$, i.e.,
  \begin{equation}
    [\lambda/z]_i \times 2 = \frac {2} {z}\lambda - \frac{c}{\sqrt{y}},\label{equ:instaLinSqrtY}
  \end{equation}
  when plugged into Eq.~(\ref{equ:lambdaZi}).
\end{subequations}
Similarly, when using the scaling variable $y$, the exponential correction ansatz
\begin{subequations}
  \begin{equation}
    C_\text{ag} (y) \simeq a y^{-\lambda/z} \exp \left(-\frac{c}{y}\right)\label{equ:CagFitExpY}
  \end{equation}
  yields
  \begin{equation}
    [\lambda/z]_i \times 2 = \frac {2} {z}\lambda - \frac{2c}{y}.\label{equ:instaLinY}
  \end{equation}
\end{subequations}

Similar to the fitting exercise in Sec.~\ref{sec:results}, we have carried out fits of $C_\text{ag}$ using the ansaetze Eqs.~(\ref{equ:CagFitExpSqrtY}) and (\ref{equ:CagFitExpY}) on the same fitting range $y\in[76,290]$, giving for the $1/\sqrt{y}$ ansatz in Eq.~(\ref{equ:CagFitExpSqrtY}) the parameters
\begin{equation}
  2\lambda/z = 1.73(13), \, a = 13.0(5.6), \, c^2 = 24(13), \label{equ:fittingParamsExpSqrtY}
\end{equation}
and for the $1/y$ ansatz [Eq.~(\ref{equ:CagFitExpY})]
\begin{equation}
  2\lambda/z = 1.527(73), \, a = 5.7(1.2), \, c = 14.0(3.7). \label{equ:fittingParamsExpLinY}
\end{equation}
As in this case, the instantaneous exponent is just a linear function we carry out fits using the numerical data for the instantaneous exponent as well. Here, we find for the $1/\sqrt{y}$ ansatz [Eq.~(\ref{equ:instaLinSqrtY})]
\begin{equation}
  2\lambda/z = 1.74(12), \, c^2 = 24(11), \label{equ:fittingParamsInstSqrtY}
\end{equation}
and for the $1/y$ ansatz [Eq.~(\ref{equ:instaLinY})]
\begin{equation}
  2\lambda/z = 1.521(65), \, c = 13.5(3.0), \label{equ:fittingParamsInstLinY}
\end{equation}
which is in very good agreement with the values obtained using the exponential fitting ansaetze for $C_\text{ag}$. 
Note that the statistical errors of the fitting parameters for the fits of the instantaneous exponent are even slightly smaller than the ones of $C_\text{ag}$. Initially, one might expect the Jackknife errors to be larger for the fits carried out on the instantaneous exponent as it is subject to various subtleties mentioned above. We suspect that this effect is reduced by the fact that only two fitting parameters are required instead of otherwise three. However, this is a mere numerical observation and we make no claim on either fitting approach (on the autocorrelation function or the one on the instantaneous exponent) being better than the other.

While the fitting parameters of the exponential $1/y$ fit in Eq.~(\ref{equ:fittingParamsExpLinY}) are fully consistent with those in Eq.~(\ref{equ:fittingParamsLin}), the statistical errors of the exponential $1/\sqrt{y}$ fit in Eq.~(\ref{equ:fittingParamsExpSqrtY}) turn out to be much larger than those in Eq.~(\ref{equ:fittingParamsSqrt}) and the estimate for $2\lambda/z$ shifts to a considerably higher value, albeit still compatible within error bars. This would increase our final estimate for $\lambda$ but due to the comparatively large uncertainties, we have not taken this fit into account.

\end{document}